\documentclass[aps,prl,reprint,superscriptaddress]{revtex4-2}

\usepackage{graphicx}
\usepackage{physics}
\usepackage{xcolor}
\begin{document}

%\preprint{}

\title{Observation of Photoinduced Terahertz Gain in GaAs Quantum Wells: Evidence for Radiative Two-Exciton-to-Biexciton Scattering} 

\normalsize

\author{Xinwei~Li}
\thanks{These authors contributed equally.}
\affiliation{Department of Electrical and Computer Engineering, Rice University, Houston, Texas 77005, USA}

\author{Katsumasa~Yoshioka}
\thanks{These authors contributed equally.}
\affiliation{Department of Physics, Graduate School of Engineering Science, Yokohama National University, Yokohama 240-8501, Japan}

\author{Qi~Zhang}
\affiliation{School of Physics, Nanjing University, Nanjing, 210093, China}

\author{Nicolas~Marquez~Peraca}
\affiliation{Department of Physics and Astronomy, Rice University, Houston, Texas 77005, USA}

\author{Fumiya~Katsutani}
\affiliation{Department of Electrical and Computer Engineering, Rice University, Houston, Texas 77005, USA}

\author{Weilu~Gao}
\affiliation{Department of Electrical and Computer Engineering, Rice University, Houston, Texas 77005, USA}

\author{G.~Timothy~Noe~II}
\affiliation{Department of Electrical and Computer Engineering, Rice University, Houston, Texas 77005, USA}

\author{John~D.~Watson}
\author{Michael~J.~Manfra}
\affiliation{Department of Physics and Astronomy, Purdue University, West Lafayette, Indiana 47907, USA}

\author{Ikufumi~Katayama}
%\email{katayama-ikufumi-bm@ynu.ac.jp}
\affiliation{Department of Physics, Graduate School of Engineering Science, Yokohama National University, Yokohama 240-8501, Japan}

\author{Jun~Takeda}
%\email{jun@ynu.ac.jp}
\affiliation{Department of Physics, Graduate School of Engineering Science, Yokohama National University, Yokohama 240-8501, Japan}

\author{Junichiro~Kono}
\email{kono@rice.edu}
\affiliation{Department of Electrical and Computer Engineering, Rice University, Houston, Texas 77005, USA}
\affiliation{Department of Physics and Astronomy, Rice University, Houston, Texas 77005, USA}
\affiliation{Department of Material Science and NanoEngineering, Rice University, Houston, Texas 77005, USA}

\date{\today}

\begin{abstract}
We have observed photoinduced negative optical conductivity, or gain, in the terahertz frequency range in a GaAs multiple-quantum-well structure in a strong perpendicular magnetic field at low temperatures.  The gain is narrow-band: it appears as a sharp peak (linewidth $<$0.45\,meV) whose frequency shifts with applied magnetic field. The gain has a circular-polarization selection rule: a strong line is observed for hole-cyclotron-resonance-active polarization. Furthermore, the gain appears only when the exciton $1s$ state is populated, which rules out intraexcitonic transitions to be its origin. Based on these observations, we propose a possible process in which the stimulated emission of a terahertz photon occurs while two free excitons scatter into one biexciton in an energy and angular-momentum conserving manner.
\end{abstract}

%\keywords{}


\maketitle

An ensemble of correlated electron-hole ($e$-$h$) pairs, or excitons, in an insulating solid in an external magnetic field ($B$), or magnetoexcitons, provides a highly tunable nonequilibrium system in which to study quantum many-body phenomena \cite{CongetAl18InBook}. The internal states of an exciton, which are analogous to the states of a hydrogen atom, can be probed by intraexciton transitions (i.e., the excitonic Lyman or Balmer series) using terahertz (THz) radiation (meV in photon energy) \cite{GershenzonetAl76JETP,LabrieetAl88PRL,HoddgeetAl90PRB,GroeneveldGrischkowsky94JOSAB,CerneetAl96PRL,SalibetAl96PRL,KonoetAl97PRL,KaindletAl03Nature,HuberetAl05PRB,HuberetAl06PRL,Lloyd-HughesetAl08PRB,SuzukiShimano09PRL,KaindletAl09PRB,SuzukiShimano12PRL,RiceetAl13PRL,BhattacharyyaetAl14PRB,LuoetAl15PRL,PollmannetAl15NM,ZhangetAl16PRL,LuoetAl17NC,LuoetAl19PRM}. These transitions evolve as a function of $B$, exhibiting shifts and splittings~\cite{CongetAl18InBook,Lloyd-HughesetAl08PRB,BhattacharyyaetAl14PRB,SekiguchiShimano15PRB,ZhangetAl16PRL}, which provide insight into dark states that are not observable in interband optical experiments. It has been recently shown, by probing the 1$s$$-$2$p_-$ transition~\cite{ZhangetAl16PRL}, that two-dimensional (2D) magnetoexcitons in high $B$ are extremely stable against an excitonic Mott transition~\cite{MacDonaldRezayi90PRB,DzyubenkoLozovik91JPA,ApalkonRashba91JETP,MacDonaldetAl92PRL,YoonetAl97SSC,RashbaetAl00SSC,KimetAl13PRB}, which would ordinarily transform the system from an insulating and bosonic exciton gas into a metallic and fermionic $e$-$h$ plasma at high pair densities. Such an observation is consistent with the hidden symmetry of 2D magnetoexcitons~\cite{MacDonaldRezayi90PRB,DzyubenkoLozovik91JPA,ApalkonRashba91JETP}, which prevents density-driven exciton dissociation. Further THz spectroscopy studies are needed to develop a microscopic understanding of many-body interactions within the 2D magnetoexciton system.

Here, we report an unexpected phenomenon observed in a 2D magnetoexciton system. We performed time-resolved and circular-polarization-resolved THz magnetospectroscopy measurements on photoexcited GaAs quantum wells (QWs) in $B$, and observed a spectroscopic feature that showed a \emph{negative} value for the real part of the optical conductivity, indicating THz gain. The gain feature had a center frequency tunable by $B$, appeared as a strong line in the hole-cyclotron-resonance-active ($h$CRA) circularly polarized mode, and disappeared at temperatures ($T$) higher than 10~K. We further found that the \emph{upper} state of population inversion (USPI) was the 1$s$ exciton state, which rules out any intraexcitonic transition as the gain transition. Based on these observations, we propose the following possible scenario to explain the THz gain: the lower state of population inversion (LSPI) is the biexciton ground state, implying that the scattering of two free 1$s$ excitons into one biexciton caused THz stimulated emission. This process is complementary to the biexciton-exciton population inversion observed in GaAs quantum wires, which resulted in gain in the near-infrared (NIR) range~\cite{HayamizuetAl07PRL}.

\begin{figure}[t!]
	\begin{center}
		\includegraphics[width=\linewidth]{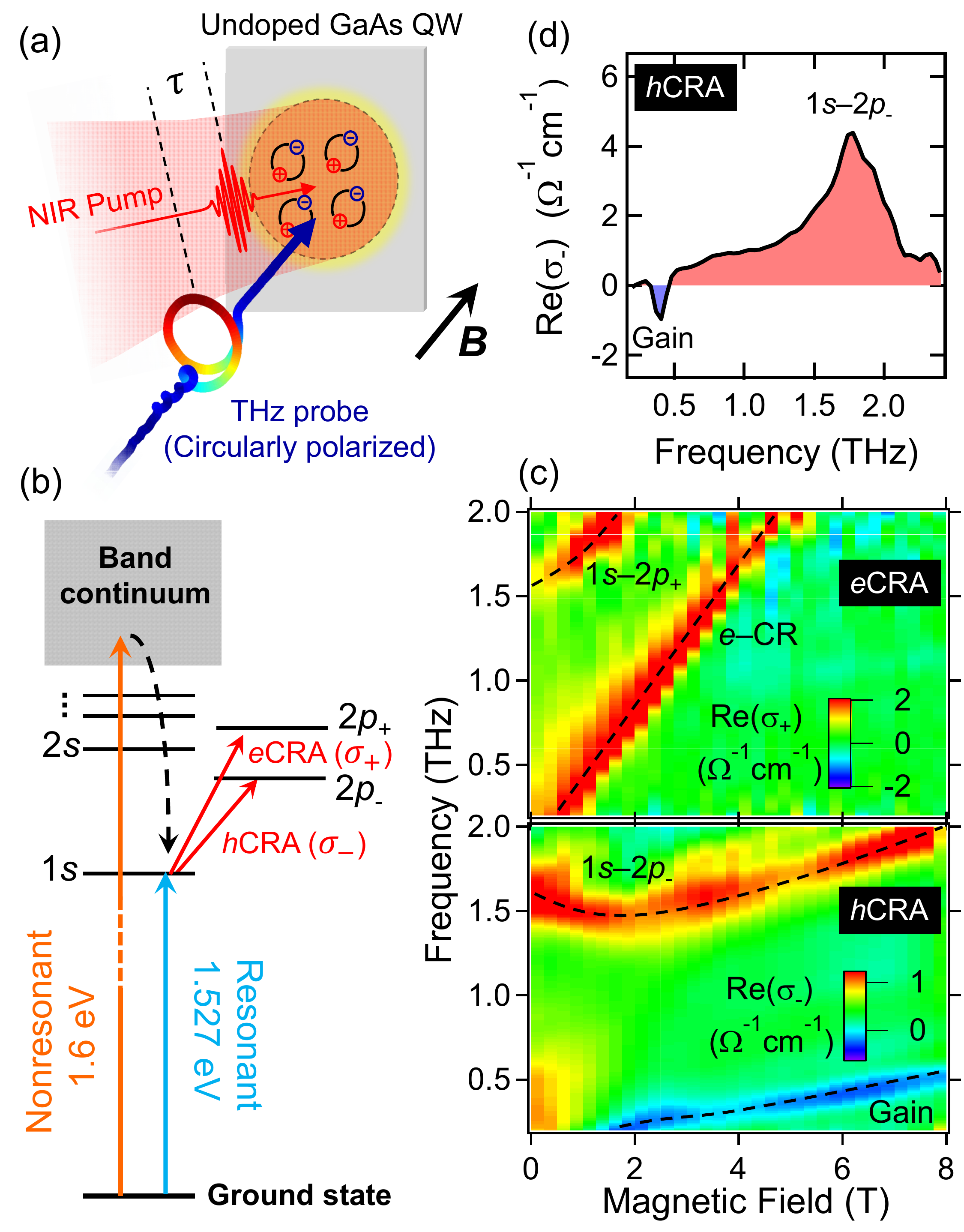}
		\caption{\small (a)~Schematic diagram of the experimental geometry. (b)~Energy-level diagram showing the transitions used for the optical pump and circularly polarized THz probe for both resonant and nonresonant pumping experiments.  (c)~Experimental THz magnetoconductivity maps for nonresonant pumping with $T=2$~K, $E=1.60$~eV,  $F=400$~$\text{nJ/cm}^2$, and $\tau=900$~ps. The upper and lower panels in (c) map $\text{Re}(\sigma_+)$ and $\text{Re}(\sigma_-)$, respectively.  Black dashed lines are guides to the eye. (d)~$\text{Re}(\sigma_-)$ spectrum obtained in a resonant pumping experiment with $E=1.527$~eV, $T=4$~K, $B=7$~T, $F=250$~$\text{nJ/cm}^2$, and $\tau=300$~ps.}
		\label{figure1}
	\end{center}
\end{figure}

Our experimental scheme is shown in Fig.\,\ref{figure1}(a). We performed optical-pump-THz-probe magnetospectroscopy experiments on an undoped GaAs multiple-QW sample grown by molecular beam epitaxy. The QW sample contained 15 periods of alternating layers of 20-nm-wide GaAs wells and 20-nm-wide Al$_{0.3}$Ga$_{0.7}$As barriers. To ensure that photoinduced signals originated exclusively from the QW active region rather than the bulk GaAs buffer layer and substrate, we performed a substrate-removal procedure by wet chemical etching, and transferred the QW active region onto a transparent sapphire substrate. 

Our laser system was a Ti:sapphire regenerative amplifier (1~kHz, 150~fs, 775~nm, Clark-MXR, inc.), which generated and detected THz probe pulses with ZnTe crystals and also fed an optical parametric amplifier system for NIR pump generation. The NIR pump %had a fluence of $F$; the pump 
photon energy, $E$, was tunable from 1.52~eV to 1.60~eV.  The spectral width of the pump, $\Delta E$, was around 3~meV after spectral filtering by a $4f$ pulse shaper, which was sharp enough to selectively pump various band-edge exciton states or the band continuum. 

The THz probe pulse arrived at the sample at time delay $\tau$ after the NIR pump pulse excited the sample.  THz probe pulses covered a frequency range of 0.2--2.5~THz and were circularly polarized by using an achromatic THz quarter-wave plate~\cite{KawadaetAl14OL,LietAl18NP}. In $B$, due to the opposite signs of Lorentz forces, electron cyclotron resonance (CR) and hole CR respond to circularly polarized THz light with opposite helicities. We denote the two circular polarizations as the electron-CR-active ($e$CRA) mode and the $h$CRA mode. THz conductivities for the $e$CRA and $h$CRA modes, denoted as $\sigma_+$ and $\sigma_-$, respectively, were measured.

Figure\,\ref{figure1}(b) displays an energy-level diagram showing the pump and probe transitions used in both nonresonant and resonant pumping experiments.  Figure\,\ref{figure1}(c) shows an experimental THz magnetoconductivity map obtained in the nonresonant pumping case.  Here, the pump fluence ($F$) was 400~$\text{nJ/cm}^2$, and other experimental conditions were $T=2$~K, $E=1.60$~eV, and $\tau=900$~ps. The pump photon energy was 75~meV above the heavy-hole exciton $1s$ energy at 0~T; note that we will focus on heavy-hole excitons in this article.  Therefore, the pump created a hot $e$-$h$ plasma in the continuum when it arrived at the sample (orange arrow). Carrier cooling and exciton formation ensued in the next 900~ps (black dashed arrow), so that by the time the THz probe arrived at the sample, a subset of the $e$-$h$ pairs were occupying the $1s$ exciton state.  The $2p$ state split into two ($2p_+$ and $2p_-$) in $B$ due to the nonzero orbital angular momentum. The THz probe light with $e$CRA ($h$CRA) circular polarization induced the 1$s$--2$p_+$ (1$s$--2$p_-$) intraexcitonic transition, as shown by the red solid arrows. 

The upper (lower) panel of Fig.\,\ref{figure1}(c) maps $\text{Re}(\sigma_+)$ for $e$CRA polarization ($\text{Re}(\sigma_-)$ for $h$CRA polarization) as a function of frequency and magnetic field. Two absorption lines with $\text{Re}(\sigma_+)>0$ can be observed in the upper panel. The line that starts from 1.6~THz at $B=0$~T and increases in frequency with $B$ is the $1s$--$2p_+$ transition. The line that starts from zero frequency at 0~T and increases linearly with $B$ is due to the CR of free electrons ($e$-CR), which are nonresonantly created and have not relaxed to the 1$s$ exciton state.

In the lower panel of Fig.\,\ref{figure1}(c) showing $\text{Re}(\sigma_-)$, the $e$-CR and $1s$--$2p_+$ transitions are not observed, because their polarization selection rules are not satisfied. In contrast, the $1s$--$2p_-$ transition, which is allowed for $h$CRA-polarized radiation, appears; its frequency first decreases and then increases with $B$ due to the combined effects of Zeeman splitting and diamagnetic shift~\cite{CongetAl18InBook}. Most notably, there is a narrow line exhibiting $\text{Re}(\sigma_-)<0$ on the low frequency side of the colormap, indicating gain. The gain peak frequency increases with $B$ and only appears in the $h$CRA polarization mode. The gain feature does not appear in measurements on bulk GaAs crystals under the same experimental conditions~\cite{SM}.
 
Figure\,\ref{figure1}(d) shows a $\text{Re}(\sigma_-)$ spectrum (i.e., for $h$CRA polarization) taken under the following conditions: $E=1.527$~eV, $T=4$~K, $B=7$~T, $F=250$~$\text{nJ/cm}^2$, and $\tau=300$~ps.  The pump photon energy $E$ was resonant with the 1$s$ exciton energy (blue arrow in Fig.\,\ref{figure1}(b)), and hence, 1$s$ excitons were resonantly created without exciting free carriers. Both the gain feature and the $1s$--$2p_-$ absorption feature clearly appear in the spectrum with opposite signs. 

\begin{figure}[t!]
	\begin{center}
		\includegraphics[width=\linewidth]{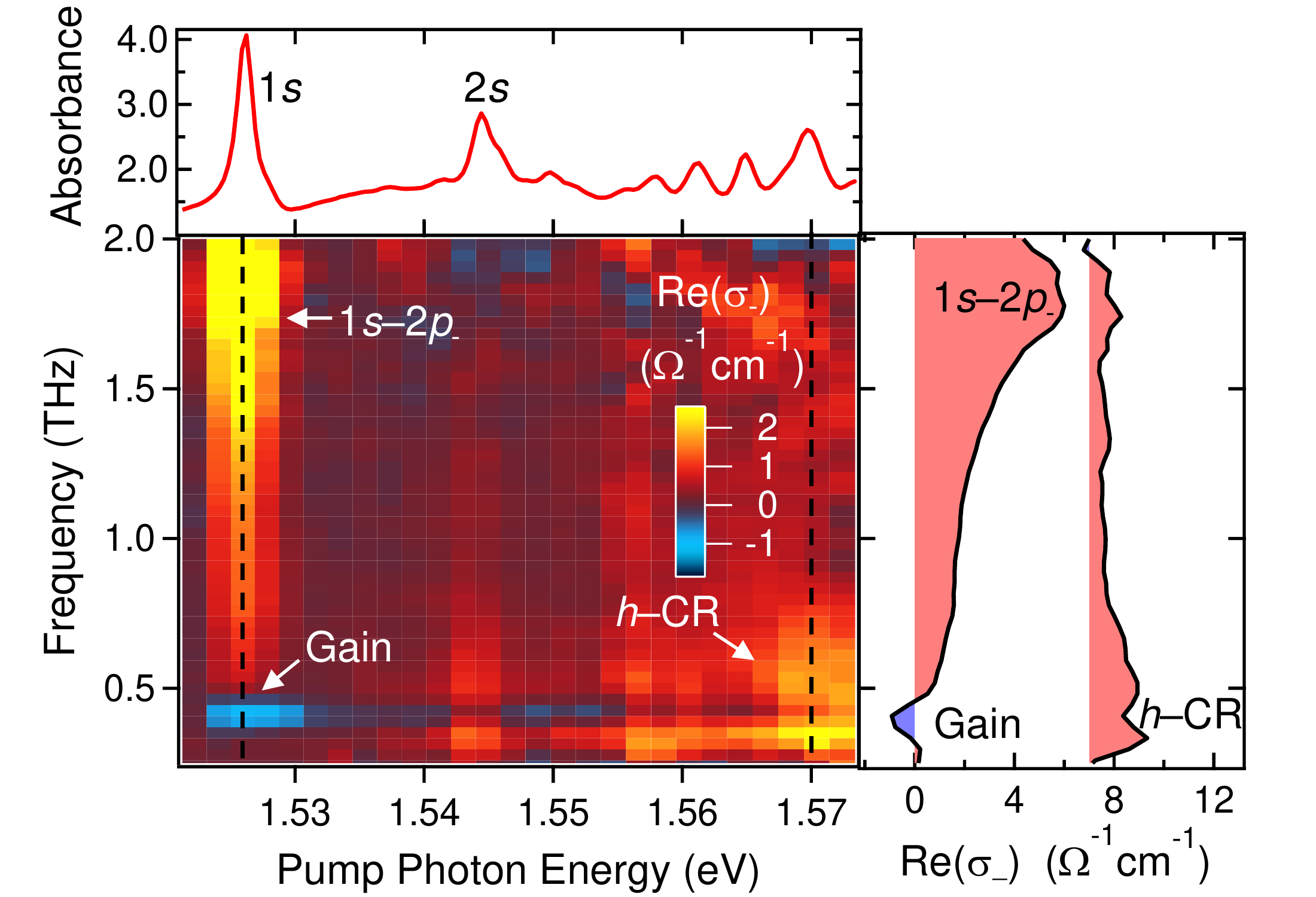}
		\caption{\small $\text{Re}(\sigma_-)$ map versus THz frequency and pump photon energy at $T=4$~K, $B=7$~T, $F=250$~$\text{nJ/cm}^2$, and $\tau=15$~ps. The top panel shows a linear absorbance spectrum for the QW at the same $T$ and $B$. The right panel shows two THz spectra at $E=1.527$~eV and $E=1.570$~eV, corresponding to the two vertical cuts marked by the black dashed lines in the map.}
		\label{figure2}
	\end{center}
\end{figure}

The appearance of gain in the resonant pumping experiment precludes the possibility that the 1$s$ state is the LSPI.  Namely, THz gain observed here is \emph{not} due to an intraexciton transition~\cite{KiraKoch04PRL,HuberetAl06PRL}.  To obtain more information on the possible USPI, we performed pump photon energy dependence measurements. Figure\,\ref{figure2} shows $\text{Re}(\sigma_-)$ spectra obtained with various pump photon energies while we kept $T=4$~K, $B=7$~T, $F=250$~$\text{nJ/cm}^2$, and $\tau=15$~ps; a linear absorbance spectrum of the QW at the same $T$ and $B$ is shown on the top, displaying which state is pumped. The value of $\tau=15$~ps is shorter than the characteristic intraband carrier cooling time. Pronounced features due to the $1s$--$2p_-$ transition and the gain feature again appear only when the 1$s$ state is resonantly pumped, that is, $E=1.527$~eV. 

Pumping with higher $E$ photons creates excitons in excited states, such as $2s$ and $3s$; a Mott transition occurs for these excited excitons, and the resulting free heavy-hole CR absorption becomes bright (the absorption feature marked as ``$h$-CR" in the colormap). The frequencies of the gain feature and the $h$-CR absorption are close, as shown in the two THz spectra on the right, corresponding to the two vertical cuts at $E=1.527$~eV and $E=1.570$~eV in the color map. However, when the gain feature appears in the resonant pumping case, the $h$-CR absorption is clearly absent, reflecting the absence of a Mott transition for 1$s$ excitons.

\begin{figure}[t!]
	\begin{center}
		\includegraphics[width=\linewidth]{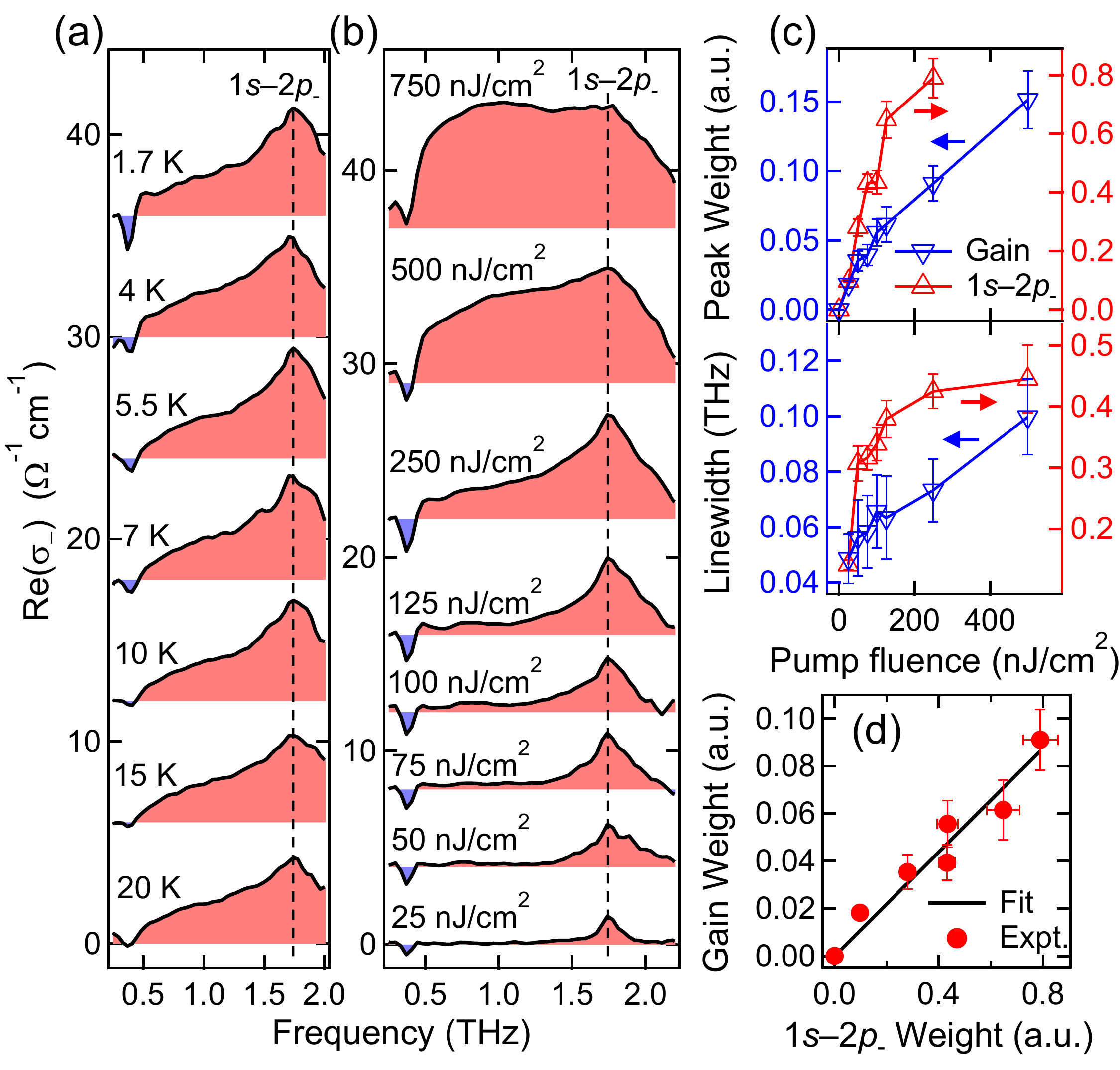}
		\caption{\small (a)~Temperature dependence of $\text{Re}(\sigma_-)$ spectra obtained with $E=1.527$~eV (resonant pumping), $B=7$~T, $F=250$~$\text{nJ/cm}^2$, and $\tau=15$~ps. (b)~Pump fluence dependence of $\text{Re}(\sigma_-)$ spectra obtained with $E=1.527$~eV (resonant pumping), $T=2$~K, $B=7$~T, and $\tau=15$~ps. (c)~Integrated peak weights and linewidths of the gain and the $1s$--$2p_-$ transition peaks versus pump fluence. (d)~Gain weight is proportional to the $1s$--$2p_-$ transition weight.}
		\label{figure3}
	\end{center}
\end{figure}

Figure\,\ref{figure3}(a) shows $\text{Re}(\sigma_-)$ spectra at different temperatures.  We kept the other experimental parameters at $E=1.527$~eV (resonant pumping), $B=7$~T, $F=250$~$\text{nJ/cm}^2$, and $\tau=15$~ps. There is a clear trend that the amplitude of the $\text{Re}(\sigma_-)<0$ dip due to the gain feature decreases with increasing $T$.  At around 10~K, the gain feature disappears.

Furthermore, we performed pump fluence dependent measurements for quantifying the maximally achievable THz gain under NIR pumping with increasing intensity. Figure\,\ref{figure3}(b) shows $\text{Re}(\sigma_-)$ spectra at various pump fluences from $F=25$~$\text{nJ/cm}^2$ to $750$~$\text{nJ/cm}^2$ while we kept $E=1.527$~eV (resonant pumping), $T=2$~K, $B=7$~T, and $\tau=15$~ps. We fit these spectra with a model in which $\text{Re}(\sigma_-)$ consists of a Lorentzian dip at 0.38~THz (the gain feature), a Lorentzian peak at 1.75~THz (the $1s$--$2p_-$ absorption), and a broadband background~\cite{SM}. Integrated weights and linewidths of the gain peak and the $1s$--$2p_-$ absorption peak are extracted from the fit, and plotted versus $F$ in Fig.\,\ref{figure3}(c). 

The gain feature appears at the smallest $F$ and grows with increasing $F$ until $250$~$\text{nJ/cm}^2$. Above $250$~$\text{nJ/cm}^2$, the gain feature saturates and eventually merges with the strong $\text{Re}(\sigma_-)>0$ background; the region where $\text{Re}(\sigma_-)<0$ decreases and eventually disappears at $750$~$\text{nJ/cm}^2$. The maximum gain coefficient achieved is 0.5~cm$^{-1}$, which appears at the center of the narrow gain band at $250$~$\text{nJ/cm}^2$ (the photoexcited $e$-$h$ pair density is $7.95\times10^9$ $\text{cm}^{-2}$ per well). 
The weight of the $1s$--$2p_-$ transition also shows saturation. When we plot the weight of the gain versus that of the $1s$--$2p_-$ transition, proportionality is seen; see Fig.\,\ref{figure3}(d). This relation suggests that the exciton 1$s$ state is the USPI, because the gain coefficient should be proportional to the inverted population difference. This conclusion agrees with the pump photon energy dependence study in Fig.\,\ref{figure2}. The observed proportionality of the peak weights, along with several other observations, allows us to rule out any impurity-related mechanisms for the gain~\cite{SM}. The linewidths of both features increase with $F$ (Fig.\,\ref{figure3}(c)), possibly due to the increasing scattering rate with increasing exciton density.

\begin{figure}[t!]
	\begin{center}
		\includegraphics[width=\linewidth]{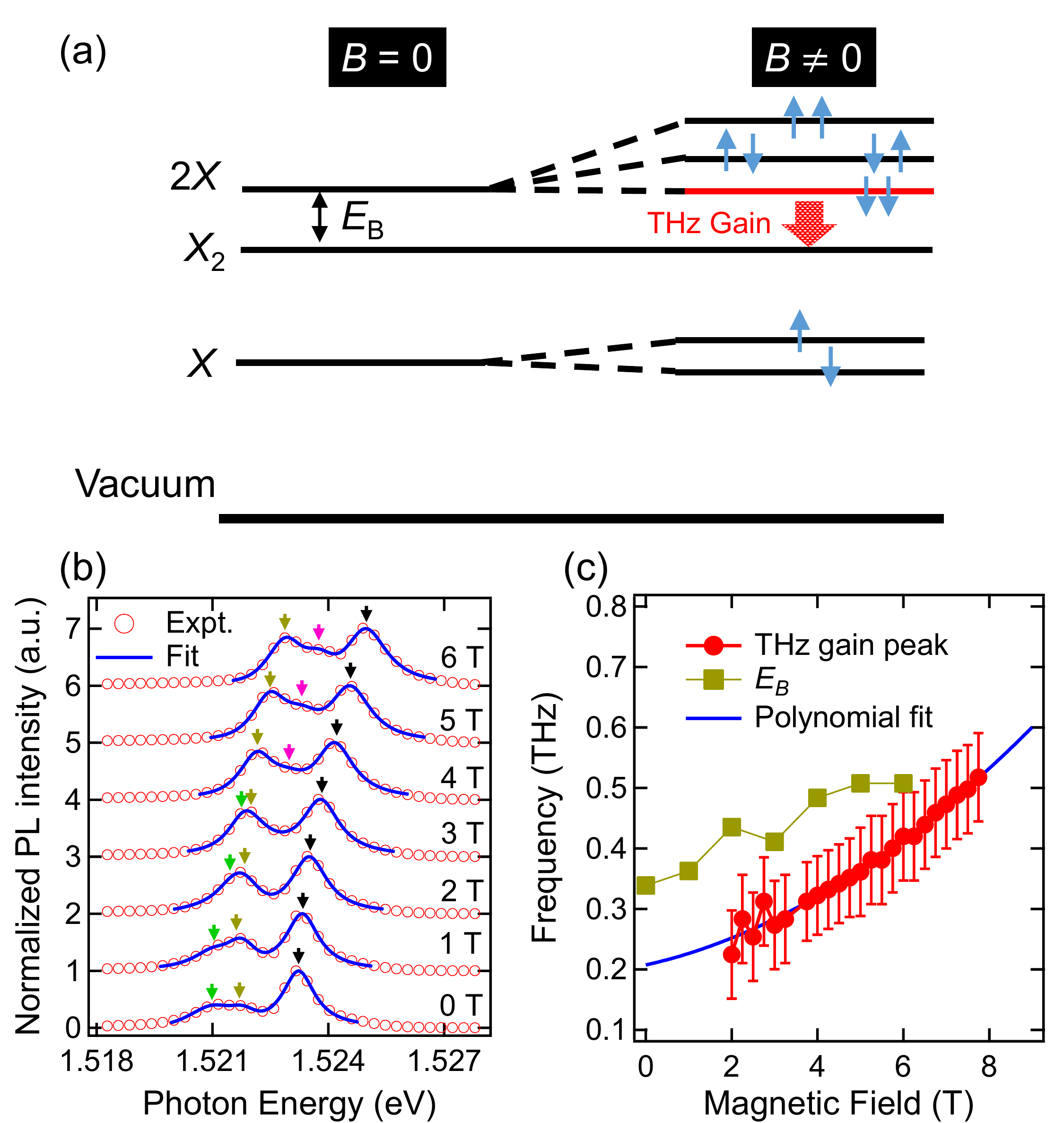}
		\caption{\small (a)~Proposed transition scheme for the gain feature. (b)~Magnetic field dependence of PL spectra; both experimental data and peak fits are plotted. All curves are normalized with respect to the free exciton peak intensity and vertically offset for clarity. Black, green, blue, and pink arrows mark the positions of the free exciton, charged exciton, biexciton, and unknown peaks, respectively.  (c)~Biexciton binding energy estimated from PL measurements plotted together with the THz gain frequency versus magnetic field.}
		\label{figure4}
	\end{center}
\end{figure}

We succinctly summarize the essential experimental facts here.  The gain feature is narrow-band, is tunable with $B$, appears as a strong line for $h$CRA circular polarized radiation, disappears at $T>10$~K, and requires finite 1$s$ state population. The last point shows that the 1$s$ exciton state is the USPI, which narrows down the choices for the LSPI because the 1$s$ state is the ground state of excitons.  Any state that is energetically lower and has an excitonic origin must be from a more complicated $e$-$h$ complex where an exciton binds to an additional object so that the total energy of the bound $e$-$h$ complex is lower than its constituents in the state where they are free to move about. Following this logic, we discuss below a possible physical origin of the LSPI. 

We propose that the LSPI is the biexciton ground state, and therefore, the THz gain transition is from the two-exciton state (denoted as $2X$), describing two free 1$s$ excitons, to the biexciton ground state (denoted as $X_2$).  As shown in the schematic diagram in Fig.\,\ref{figure4}(a), when the NIR pump resonantly creates excitons ($X$), the $2X$ state is automatically populated. At $B\neq0$, the $2X$ state shifts, and splits into a multiplet due to several choices of the spin alignment of the two excitons. The NIR pump populates the multiplet evenly, but the exciton spin slowly relaxes \cite{HarleyetAL96PRB} during the time interval between the pump and probe. For a long enough delay time $\tau$, the lowest state of the multiplet, $\ket{\downarrow\downarrow}$, is most populated when the THz probe arrives as the thermal energy is smaller than the spin splitting energy. The gain transition observed in the $h$CRA mode is from $\ket{\downarrow\downarrow}$ to $X_2$. Whether or not such a transition is able to cause stimulated emission, or even radiatively allowed, has not been discussed or investigated in previous studies.  Note that this process has some similarity to the $X_2$--$X$ population inversion observed in quantum wires~\cite{HayamizuetAl07PRL}, which, however, required the $X_2$ density to be higher than the $X$ density and the gain appeared in the NIR. It is also interesting to compare our scheme to the THz luminescence and gain reported in GaAs QWs and silicon samples that are intentionally doped with shallow impurities, where the major emission/gain features are much higher in energy than our gain feature and are ascribed to intra-impurity-center transitions~\cite{ZhukavinetAL07JAP,ZakharinetAL10APL,MakhovetAL19JLum,MakhovetAL19JAP}. Below, we provide additional supporting evidence that our proposed scenario can explain our several experimental observations, assuming that this radiative transition is allowed.

First, based on this scenario, the energy of the THz gain should be the biexciton binding energy, $E_\text{B}$, which can be measured independently by photoluminescence (PL) spectroscopy. Figure~\ref{figure4}(b) shows PL spectra at $T=4$~K in $B$ up to 6~T.  There are peaks due to biexcitons (blue arrows) and charged excitons (green arrows) on the lower energy side of the free exciton 1$s$ peak (black arrows) at $T<10$~K, which is the $T$ range in which the THz gain was observed. See Ref.~\cite{SM} for details of PL peak assignments.  $E_\text{B}$ measured from Fig.\,\ref{figure4}(b) is plotted against $B$ in Fig.\,\ref{figure4}(c) together with the center frequency of the observed THz gain feature. $E_\text{B}$ presents a reasonable match to the gain transition frequency, but it overestimates the curve, possibly because it is calculated by the energy difference between the biexciton peak and the \emph{center} of the free exciton peak in the PL spectra, while the gain is from the lower edge of the $2X$ multiplet. The trend that $E_\text{B}$ increases with $B$ matches the behavior of the magnetic field dependence of the THz gain.

Second, we examine how our proposed scheme can explain the polarization selection rule of the THz gain. It is known that the LSPI, namely, the $X_2$ state, is a spin singlet state with zero total angular momentum. Therefore, the polarization selection rule of the gain transition is determined by the spin angular momentum carried by the $\ket{\downarrow\downarrow}$ state in Fig.\,\ref{figure4}(a). Both the sign of the exciton $g$ factor~\cite{SnellingetAl92PRB} and hole-angular-momentum mixing~\cite{TraynoretAl97PRB} due to valence band hybridization are important. We took into consideration both factors and confirmed that the wavefunction of the $\ket{\downarrow\downarrow}$ state possesses a component that supports THz gain for the $h$CRA polarization~\cite{SM}. On the other hand, the $\ket{\uparrow\uparrow}$ state, if populated, would support THz gain for the $e$CRA polarization channel, and the gain frequency should be higher than the $h$CRA polarized gain. However, an $e$CRA circularly polarized gain line cannot be distinguished in Fig.\,\ref{figure1}(c). A possible reason is that the exciton spin population is mostly relaxed during the long delay time ($\tau=900$~ps). It would be interesting to perform experiments in the future to confirm if an $e$CRA-polarized gain appears at short delay times in the resonant pumping condition in $B$ at low $T$. Tracking the full time delay dependence of this feature would also enable direct measurement of the $B$-dependent exciton spin relaxation time by monitoring how the gain spectral weight transfers from the $e$CRA polarization to the $h$CRA polarization as population transfers from the $\ket{\uparrow\uparrow}$ state to the $\ket{\downarrow\downarrow}$ state; analysis of preliminary time delay dependence data for the $h$CRA polarized gain suggests that the spectral weight transfer indeed occurs~\cite{SM}. Another fascinating experiment utilizing the polarization selection rule is ultrafast switching of gain by controlling the optical orientation. A circularly polarized NIR pump can selectively populate the $\ket{\uparrow\uparrow}$ state or the $\ket{\downarrow\downarrow}$ state so the THz gain can be switched between the $e$CRA and $h$CRA channels.

In summary, by performing optical-pump-THz-probe experiments on undoped GaAs quantum wells at low temperatures in a strong perpendicular magnetic field, we observed a narrow-band, circularly polarized THz gain feature whose frequency shifts with applied magnetic field. From systematic experiments as a function of magnetic field, temperature, pump fluence, pump photon energy, and probe polarization, we confirmed that the upper state of population inversion is the 1$s$ exciton state.  We described a possible scenario of the biexciton ground state being the lower state of population inversion and proposed a specific transition scheme for the gain to appear.
Our observations not only provide new insight into the physics of many-exciton states and radiative transitions in a complex 2D $e$-$h$ system in a strong magnetic field but also open up possibilities for developing tunable THz lasers based on 2D magnetoexcitons.

\begin{acknowledgments}
We thank Cun-Zheng Ning, Peter Littlewood, Alexey Belyanin, Yongrui Wang, and Mackillo Kira for useful discussions. We thank Yoichi Kawada, Hironori Takahashi and Hamamatsu Photonics K.K. for fabricating the achromatic THz quarter-wave plate. I.K.\ and J.T.\ acknowledge support from the Ministry of Education, Culture, Sports, Science and Technology (MEXT) /Japan Society for the Promotion of Science (JSPS) through KAKENHI Grant Nos.\ 17H06124 and 18H04288.
\end{acknowledgments}

%\bibliography{jun}

\begin{thebibliography}{42}%
\makeatletter
\providecommand \@ifxundefined [1]{%
 \@ifx{#1\undefined}
}%
\providecommand \@ifnum [1]{%
 \ifnum #1\expandafter \@firstoftwo
 \else \expandafter \@secondoftwo
 \fi
}%
\providecommand \@ifx [1]{%
 \ifx #1\expandafter \@firstoftwo
 \else \expandafter \@secondoftwo
 \fi
}%
\providecommand \natexlab [1]{#1}%
\providecommand \enquote  [1]{``#1''}%
\providecommand \bibnamefont  [1]{#1}%
\providecommand \bibfnamefont [1]{#1}%
\providecommand \citenamefont [1]{#1}%
\providecommand \href@noop [0]{\@secondoftwo}%
\providecommand \href [0]{\begingroup \@sanitize@url \@href}%
\providecommand \@href[1]{\@@startlink{#1}\@@href}%
\providecommand \@@href[1]{\endgroup#1\@@endlink}%
\providecommand \@sanitize@url [0]{\catcode `\\12\catcode `\$12\catcode
  `\&12\catcode `\#12\catcode `\^12\catcode `\_12\catcode `\%12\relax}%
\providecommand \@@startlink[1]{}%
\providecommand \@@endlink[0]{}%
\providecommand \url  [0]{\begingroup\@sanitize@url \@url }%
\providecommand \@url [1]{\endgroup\@href {#1}{\urlprefix }}%
\providecommand \urlprefix  [0]{URL }%
\providecommand \Eprint [0]{\href }%
\providecommand \doibase [0]{https://doi.org/}%
\providecommand \selectlanguage [0]{\@gobble}%
\providecommand \bibinfo  [0]{\@secondoftwo}%
\providecommand \bibfield  [0]{\@secondoftwo}%
\providecommand \translation [1]{[#1]}%
\providecommand \BibitemOpen [0]{}%
\providecommand \bibitemStop [0]{}%
\providecommand \bibitemNoStop [0]{.\EOS\space}%
\providecommand \EOS [0]{\spacefactor3000\relax}%
\providecommand \BibitemShut  [1]{\csname bibitem#1\endcsname}%
\let\auto@bib@innerbib\@empty
%</preamble>
\bibitem [{\citenamefont {Cong}\ \emph {et~al.}(2018)\citenamefont {Cong},
  \citenamefont {Noe~II},\ and\ \citenamefont {Kono}}]{CongetAl18InBook}%
  \BibitemOpen
  \bibfield  {author} {\bibinfo {author} {\bibfnamefont {K.}~\bibnamefont
  {Cong}}, \bibinfo {author} {\bibfnamefont {G.~T.}\ \bibnamefont {Noe~II}},\
  and\ \bibinfo {author} {\bibfnamefont {J.}~\bibnamefont {Kono}},\ }\bibinfo
  {title} {Excitons in magnetic fields},\ in\ \href
  {https://doi.org/https://doi.org/10.1016/B978-0-12-803581-8.09584-9} {\emph
  {\bibinfo {booktitle} {Encyclopedia of Modern Optics (Second Edition)}}}\
  (\bibinfo  {publisher} {Elsevier},\ \bibinfo {address} {Oxford},\ \bibinfo
  {year} {2018})\ pp.\ \bibinfo {pages} {63--81}\BibitemShut {NoStop}%
\bibitem [{\citenamefont {Gershenzon}\ \emph {et~al.}(1976)\citenamefont
  {Gershenzon}, \citenamefont {Gol'tsman},\ and\ \citenamefont
  {Ptitsina}}]{GershenzonetAl76JETP}%
  \BibitemOpen
  \bibfield  {author} {\bibinfo {author} {\bibfnamefont {E.~M.}\ \bibnamefont
  {Gershenzon}}, \bibinfo {author} {\bibfnamefont {G.~N.}\ \bibnamefont
  {Gol'tsman}},\ and\ \bibinfo {author} {\bibfnamefont {N.~G.}\ \bibnamefont
  {Ptitsina}},\ }\href@noop {} {\bibfield  {journal} {\bibinfo  {journal} {Sov.
  Phys. JETP}\ }\textbf {\bibinfo {volume} {43}},\ \bibinfo {pages} {116}
  (\bibinfo {year} {1976})}\BibitemShut {NoStop}%
\bibitem [{\citenamefont {Labrie}\ \emph {et~al.}(1988)\citenamefont {Labrie},
  \citenamefont {Thewalt}, \citenamefont {Booth},\ and\ \citenamefont
  {Kirczenow}}]{LabrieetAl88PRL}%
  \BibitemOpen
  \bibfield  {author} {\bibinfo {author} {\bibfnamefont {D.}~\bibnamefont
  {Labrie}}, \bibinfo {author} {\bibfnamefont {M.~L.~W.}\ \bibnamefont
  {Thewalt}}, \bibinfo {author} {\bibfnamefont {I.~J.}\ \bibnamefont {Booth}},\
  and\ \bibinfo {author} {\bibfnamefont {G.}~\bibnamefont {Kirczenow}},\
  }\href@noop {} {\bibfield  {journal} {\bibinfo  {journal} {Phys. Rev. Lett.}\
  }\textbf {\bibinfo {volume} {61}},\ \bibinfo {pages} {1882} (\bibinfo {year}
  {1988})}\BibitemShut {NoStop}%
\bibitem [{\citenamefont {Hodge}\ \emph {et~al.}(1990)\citenamefont {Hodge},
  \citenamefont {Phillips}, \citenamefont {Skolnick}, \citenamefont {Smith},
  \citenamefont {Whitehouse}, \citenamefont {Dawson},\ and\ \citenamefont
  {Foxon}}]{HoddgeetAl90PRB}%
  \BibitemOpen
  \bibfield  {author} {\bibinfo {author} {\bibfnamefont {C.~C.}\ \bibnamefont
  {Hodge}}, \bibinfo {author} {\bibfnamefont {C.~C.}\ \bibnamefont {Phillips}},
  \bibinfo {author} {\bibfnamefont {M.~S.}\ \bibnamefont {Skolnick}}, \bibinfo
  {author} {\bibfnamefont {G.~W.}\ \bibnamefont {Smith}}, \bibinfo {author}
  {\bibfnamefont {C.~R.}\ \bibnamefont {Whitehouse}}, \bibinfo {author}
  {\bibfnamefont {P.}~\bibnamefont {Dawson}},\ and\ \bibinfo {author}
  {\bibfnamefont {C.~T.}\ \bibnamefont {Foxon}},\ }\href@noop {} {\bibfield
  {journal} {\bibinfo  {journal} {Phys. Rev. B}\ }\textbf {\bibinfo {volume}
  {41}},\ \bibinfo {pages} {12319} (\bibinfo {year} {1990})}\BibitemShut
  {NoStop}%
\bibitem [{\citenamefont {Groeneveld}\ and\ \citenamefont
  {Grischkowsky}(1994)}]{GroeneveldGrischkowsky94JOSAB}%
  \BibitemOpen
  \bibfield  {author} {\bibinfo {author} {\bibfnamefont {R.~H.~M.}\
  \bibnamefont {Groeneveld}}\ and\ \bibinfo {author} {\bibfnamefont
  {D.}~\bibnamefont {Grischkowsky}},\ }\href@noop {} {\bibfield  {journal}
  {\bibinfo  {journal} {J. Opt. Soc. Am. B}\ }\textbf {\bibinfo {volume}
  {11}},\ \bibinfo {pages} {2502} (\bibinfo {year} {1994})}\BibitemShut
  {NoStop}%
\bibitem [{\citenamefont {Cerne}\ \emph {et~al.}(1996)\citenamefont {Cerne},
  \citenamefont {Kono}, \citenamefont {Sherwin}, \citenamefont {Sundaram},
  \citenamefont {Gossard},\ and\ \citenamefont {Bauer}}]{CerneetAl96PRL}%
  \BibitemOpen
  \bibfield  {author} {\bibinfo {author} {\bibfnamefont {J.}~\bibnamefont
  {Cerne}}, \bibinfo {author} {\bibfnamefont {J.}~\bibnamefont {Kono}},
  \bibinfo {author} {\bibfnamefont {M.~S.}\ \bibnamefont {Sherwin}}, \bibinfo
  {author} {\bibfnamefont {M.}~\bibnamefont {Sundaram}}, \bibinfo {author}
  {\bibfnamefont {A.~C.}\ \bibnamefont {Gossard}},\ and\ \bibinfo {author}
  {\bibfnamefont {G.~E.~W.}\ \bibnamefont {Bauer}},\ }\href@noop {} {\bibfield
  {journal} {\bibinfo  {journal} {Phys. Rev. Lett.}\ }\textbf {\bibinfo
  {volume} {77}},\ \bibinfo {pages} {1131} (\bibinfo {year}
  {1996})}\BibitemShut {NoStop}%
\bibitem [{\citenamefont {Salib}\ \emph {et~al.}(1996)\citenamefont {Salib},
  \citenamefont {Nickel}, \citenamefont {Herold}, \citenamefont {Petrou},
  \citenamefont {McCombe}, \citenamefont {Chen}, \citenamefont {Bajaj},\ and\
  \citenamefont {Schaff}}]{SalibetAl96PRL}%
  \BibitemOpen
  \bibfield  {author} {\bibinfo {author} {\bibfnamefont {M.~S.}\ \bibnamefont
  {Salib}}, \bibinfo {author} {\bibfnamefont {H.~A.}\ \bibnamefont {Nickel}},
  \bibinfo {author} {\bibfnamefont {G.~S.}\ \bibnamefont {Herold}}, \bibinfo
  {author} {\bibfnamefont {A.}~\bibnamefont {Petrou}}, \bibinfo {author}
  {\bibfnamefont {B.~D.}\ \bibnamefont {McCombe}}, \bibinfo {author}
  {\bibfnamefont {R.}~\bibnamefont {Chen}}, \bibinfo {author} {\bibfnamefont
  {K.~K.}\ \bibnamefont {Bajaj}},\ and\ \bibinfo {author} {\bibfnamefont
  {W.}~\bibnamefont {Schaff}},\ }\href@noop {} {\bibfield  {journal} {\bibinfo
  {journal} {Phys. Rev. Lett.}\ }\textbf {\bibinfo {volume} {77}},\ \bibinfo
  {pages} {1135} (\bibinfo {year} {1996})}\BibitemShut {NoStop}%
\bibitem [{\citenamefont {Kono}\ \emph {et~al.}(1997)\citenamefont {Kono},
  \citenamefont {Su}, \citenamefont {Inoshita}, \citenamefont {Noda},
  \citenamefont {Sherwin}, \citenamefont {Allen},\ and\ \citenamefont
  {Sakaki}}]{KonoetAl97PRL}%
  \BibitemOpen
  \bibfield  {author} {\bibinfo {author} {\bibfnamefont {J.}~\bibnamefont
  {Kono}}, \bibinfo {author} {\bibfnamefont {M.~Y.}\ \bibnamefont {Su}},
  \bibinfo {author} {\bibfnamefont {T.}~\bibnamefont {Inoshita}}, \bibinfo
  {author} {\bibfnamefont {T.}~\bibnamefont {Noda}}, \bibinfo {author}
  {\bibfnamefont {M.~S.}\ \bibnamefont {Sherwin}}, \bibinfo {author}
  {\bibfnamefont {S.~J.}\ \bibnamefont {Allen}},\ and\ \bibinfo {author}
  {\bibfnamefont {H.}~\bibnamefont {Sakaki}},\ }\href@noop {} {\bibfield
  {journal} {\bibinfo  {journal} {Phys. Rev. Lett.}\ }\textbf {\bibinfo
  {volume} {79}},\ \bibinfo {pages} {1758} (\bibinfo {year}
  {1997})}\BibitemShut {NoStop}%
\bibitem [{\citenamefont {Kaindl}\ \emph {et~al.}(2003)\citenamefont {Kaindl},
  \citenamefont {Carnahan}, \citenamefont {Haegele}, \citenamefont
  {Loevenich},\ and\ \citenamefont {Chemla}}]{KaindletAl03Nature}%
  \BibitemOpen
  \bibfield  {author} {\bibinfo {author} {\bibfnamefont {R.~A.}\ \bibnamefont
  {Kaindl}}, \bibinfo {author} {\bibfnamefont {M.~A.}\ \bibnamefont
  {Carnahan}}, \bibinfo {author} {\bibfnamefont {D.}~\bibnamefont {Haegele}},
  \bibinfo {author} {\bibfnamefont {R.}~\bibnamefont {Loevenich}},\ and\
  \bibinfo {author} {\bibfnamefont {D.~S.}\ \bibnamefont {Chemla}},\
  }\href@noop {} {\bibfield  {journal} {\bibinfo  {journal} {Nature}\ }\textbf
  {\bibinfo {volume} {423}},\ \bibinfo {pages} {734} (\bibinfo {year}
  {2003})}\BibitemShut {NoStop}%
\bibitem [{\citenamefont {Huber}\ \emph {et~al.}(2005)\citenamefont {Huber},
  \citenamefont {Kaindl}, \citenamefont {Schmid},\ and\ \citenamefont
  {Chemla}}]{HuberetAl05PRB}%
  \BibitemOpen
  \bibfield  {author} {\bibinfo {author} {\bibfnamefont {R.}~\bibnamefont
  {Huber}}, \bibinfo {author} {\bibfnamefont {R.~A.}\ \bibnamefont {Kaindl}},
  \bibinfo {author} {\bibfnamefont {B.~A.}\ \bibnamefont {Schmid}},\ and\
  \bibinfo {author} {\bibfnamefont {D.~S.}\ \bibnamefont {Chemla}},\
  }\href@noop {} {\bibfield  {journal} {\bibinfo  {journal} {Phys. Rev. B}\
  }\textbf {\bibinfo {volume} {72}},\ \bibinfo {pages} {161314} (\bibinfo
  {year} {2005})}\BibitemShut {NoStop}%
\bibitem [{\citenamefont {Huber}\ \emph {et~al.}(2006)\citenamefont {Huber},
  \citenamefont {Schmid}, \citenamefont {Shen}, \citenamefont {Chemla},\ and\
  \citenamefont {Kaindl}}]{HuberetAl06PRL}%
  \BibitemOpen
  \bibfield  {author} {\bibinfo {author} {\bibfnamefont {R.}~\bibnamefont
  {Huber}}, \bibinfo {author} {\bibfnamefont {B.~A.}\ \bibnamefont {Schmid}},
  \bibinfo {author} {\bibfnamefont {Y.~R.}\ \bibnamefont {Shen}}, \bibinfo
  {author} {\bibfnamefont {D.~S.}\ \bibnamefont {Chemla}},\ and\ \bibinfo
  {author} {\bibfnamefont {R.~A.}\ \bibnamefont {Kaindl}},\ }\href@noop {}
  {\bibfield  {journal} {\bibinfo  {journal} {Phys. Rev. Lett.}\ }\textbf
  {\bibinfo {volume} {96}},\ \bibinfo {pages} {017402} (\bibinfo {year}
  {2006})}\BibitemShut {NoStop}%
\bibitem [{\citenamefont {Lloyd-Hughes}\ \emph {et~al.}(2008)\citenamefont
  {Lloyd-Hughes}, \citenamefont {Beere}, \citenamefont {Ritchie},\ and\
  \citenamefont {Johnston}}]{Lloyd-HughesetAl08PRB}%
  \BibitemOpen
  \bibfield  {author} {\bibinfo {author} {\bibfnamefont {J.}~\bibnamefont
  {Lloyd-Hughes}}, \bibinfo {author} {\bibfnamefont {H.~E.}\ \bibnamefont
  {Beere}}, \bibinfo {author} {\bibfnamefont {D.~A.}\ \bibnamefont {Ritchie}},\
  and\ \bibinfo {author} {\bibfnamefont {M.~B.}\ \bibnamefont {Johnston}},\
  }\href@noop {} {\bibfield  {journal} {\bibinfo  {journal} {Phys. Rev. B}\
  }\textbf {\bibinfo {volume} {77}},\ \bibinfo {pages} {125322} (\bibinfo
  {year} {2008})}\BibitemShut {NoStop}%
\bibitem [{\citenamefont {Suzuki}\ and\ \citenamefont
  {Shimano}(2009)}]{SuzukiShimano09PRL}%
  \BibitemOpen
  \bibfield  {author} {\bibinfo {author} {\bibfnamefont {T.}~\bibnamefont
  {Suzuki}}\ and\ \bibinfo {author} {\bibfnamefont {R.}~\bibnamefont
  {Shimano}},\ }\href@noop {} {\bibfield  {journal} {\bibinfo  {journal} {Phys.
  Rev. Lett.}\ }\textbf {\bibinfo {volume} {103}},\ \bibinfo {pages} {057401}
  (\bibinfo {year} {2009})}\BibitemShut {NoStop}%
\bibitem [{\citenamefont {Kaindl}\ \emph {et~al.}(2009)\citenamefont {Kaindl},
  \citenamefont {H\"agele}, \citenamefont {Carnahan},\ and\ \citenamefont
  {Chemla}}]{KaindletAl09PRB}%
  \BibitemOpen
  \bibfield  {author} {\bibinfo {author} {\bibfnamefont {R.~A.}\ \bibnamefont
  {Kaindl}}, \bibinfo {author} {\bibfnamefont {D.}~\bibnamefont {H\"agele}},
  \bibinfo {author} {\bibfnamefont {M.~A.}\ \bibnamefont {Carnahan}},\ and\
  \bibinfo {author} {\bibfnamefont {D.~S.}\ \bibnamefont {Chemla}},\
  }\href@noop {} {\bibfield  {journal} {\bibinfo  {journal} {Phys. Rev. B}\
  }\textbf {\bibinfo {volume} {79}},\ \bibinfo {pages} {045320} (\bibinfo
  {year} {2009})}\BibitemShut {NoStop}%
\bibitem [{\citenamefont {Suzuki}\ and\ \citenamefont
  {Shimano}(2012)}]{SuzukiShimano12PRL}%
  \BibitemOpen
  \bibfield  {author} {\bibinfo {author} {\bibfnamefont {T.}~\bibnamefont
  {Suzuki}}\ and\ \bibinfo {author} {\bibfnamefont {R.}~\bibnamefont
  {Shimano}},\ }\href@noop {} {\bibfield  {journal} {\bibinfo  {journal} {Phys.
  Rev. Lett.}\ }\textbf {\bibinfo {volume} {109}},\ \bibinfo {pages} {046402}
  (\bibinfo {year} {2012})}\BibitemShut {NoStop}%
\bibitem [{\citenamefont {Rice}\ \emph {et~al.}(2013)\citenamefont {Rice},
  \citenamefont {Kono}, \citenamefont {Zybell}, \citenamefont {Winnerl},
  \citenamefont {Bhattacharyya}, \citenamefont {Schneider}, \citenamefont
  {Helm}, \citenamefont {Ewers}, \citenamefont {Chernikov}, \citenamefont
  {Koch}, \citenamefont {Chatterjee}, \citenamefont {Khitrova}, \citenamefont
  {Gibbs}, \citenamefont {Schneebeli}, \citenamefont {Breddermann},
  \citenamefont {Kira},\ and\ \citenamefont {Koch}}]{RiceetAl13PRL}%
  \BibitemOpen
  \bibfield  {author} {\bibinfo {author} {\bibfnamefont {W.~D.}\ \bibnamefont
  {Rice}}, \bibinfo {author} {\bibfnamefont {J.}~\bibnamefont {Kono}}, \bibinfo
  {author} {\bibfnamefont {S.}~\bibnamefont {Zybell}}, \bibinfo {author}
  {\bibfnamefont {S.}~\bibnamefont {Winnerl}}, \bibinfo {author} {\bibfnamefont
  {J.}~\bibnamefont {Bhattacharyya}}, \bibinfo {author} {\bibfnamefont
  {H.}~\bibnamefont {Schneider}}, \bibinfo {author} {\bibfnamefont
  {M.}~\bibnamefont {Helm}}, \bibinfo {author} {\bibfnamefont {B.}~\bibnamefont
  {Ewers}}, \bibinfo {author} {\bibfnamefont {A.}~\bibnamefont {Chernikov}},
  \bibinfo {author} {\bibfnamefont {M.}~\bibnamefont {Koch}}, \bibinfo {author}
  {\bibfnamefont {S.}~\bibnamefont {Chatterjee}}, \bibinfo {author}
  {\bibfnamefont {G.}~\bibnamefont {Khitrova}}, \bibinfo {author}
  {\bibfnamefont {H.~M.}\ \bibnamefont {Gibbs}}, \bibinfo {author}
  {\bibfnamefont {L.}~\bibnamefont {Schneebeli}}, \bibinfo {author}
  {\bibfnamefont {B.}~\bibnamefont {Breddermann}}, \bibinfo {author}
  {\bibfnamefont {M.}~\bibnamefont {Kira}},\ and\ \bibinfo {author}
  {\bibfnamefont {S.~W.}\ \bibnamefont {Koch}},\ }\href@noop {} {\bibfield
  {journal} {\bibinfo  {journal} {Phys. Rev. Lett.}\ }\textbf {\bibinfo
  {volume} {110}},\ \bibinfo {pages} {137404} (\bibinfo {year}
  {2013})}\BibitemShut {NoStop}%
\bibitem [{\citenamefont {Bhattacharyya}\ \emph {et~al.}(2014)\citenamefont
  {Bhattacharyya}, \citenamefont {Zybell}, \citenamefont {E\ss{}er},
  \citenamefont {Helm}, \citenamefont {Schneider}, \citenamefont {Schneebeli},
  \citenamefont {B\"ottge}, \citenamefont {Breddermann}, \citenamefont {Kira},
  \citenamefont {Koch}, \citenamefont {Andrews},\ and\ \citenamefont
  {Strasser}}]{BhattacharyyaetAl14PRB}%
  \BibitemOpen
  \bibfield  {author} {\bibinfo {author} {\bibfnamefont {J.}~\bibnamefont
  {Bhattacharyya}}, \bibinfo {author} {\bibfnamefont {S.}~\bibnamefont
  {Zybell}}, \bibinfo {author} {\bibfnamefont {F.}~\bibnamefont {E\ss{}er}},
  \bibinfo {author} {\bibfnamefont {M.}~\bibnamefont {Helm}}, \bibinfo {author}
  {\bibfnamefont {H.}~\bibnamefont {Schneider}}, \bibinfo {author}
  {\bibfnamefont {L.}~\bibnamefont {Schneebeli}}, \bibinfo {author}
  {\bibfnamefont {C.~N.}\ \bibnamefont {B\"ottge}}, \bibinfo {author}
  {\bibfnamefont {B.}~\bibnamefont {Breddermann}}, \bibinfo {author}
  {\bibfnamefont {M.}~\bibnamefont {Kira}}, \bibinfo {author} {\bibfnamefont
  {S.~W.}\ \bibnamefont {Koch}}, \bibinfo {author} {\bibfnamefont {A.~M.}\
  \bibnamefont {Andrews}},\ and\ \bibinfo {author} {\bibfnamefont
  {G.}~\bibnamefont {Strasser}},\ }\href@noop {} {\bibfield  {journal}
  {\bibinfo  {journal} {Phys. Rev. B}\ }\textbf {\bibinfo {volume} {89}},\
  \bibinfo {pages} {125313} (\bibinfo {year} {2014})}\BibitemShut {NoStop}%
\bibitem [{\citenamefont {Luo}\ \emph {et~al.}(2015)\citenamefont {Luo},
  \citenamefont {Chatzakis}, \citenamefont {Patz},\ and\ \citenamefont
  {Wang}}]{LuoetAl15PRL}%
  \BibitemOpen
  \bibfield  {author} {\bibinfo {author} {\bibfnamefont {L.}~\bibnamefont
  {Luo}}, \bibinfo {author} {\bibfnamefont {I.}~\bibnamefont {Chatzakis}},
  \bibinfo {author} {\bibfnamefont {A.}~\bibnamefont {Patz}},\ and\ \bibinfo
  {author} {\bibfnamefont {J.}~\bibnamefont {Wang}},\ }\href@noop {} {\bibfield
   {journal} {\bibinfo  {journal} {Phys. Rev. Lett.}\ }\textbf {\bibinfo
  {volume} {114}},\ \bibinfo {pages} {107402} (\bibinfo {year}
  {2015})}\BibitemShut {NoStop}%
\bibitem [{\citenamefont {P\"ollmann}\ \emph {et~al.}(2015)\citenamefont
  {P\"ollmann}, \citenamefont {Steinleitner}, \citenamefont {Leierseder},
  \citenamefont {Nagler}, \citenamefont {Plechinger}, \citenamefont {Porer},
  \citenamefont {Bratschitsch}, \citenamefont {Sch\"uller}, \citenamefont
  {Korn},\ and\ \citenamefont {Huber}}]{PollmannetAl15NM}%
  \BibitemOpen
  \bibfield  {author} {\bibinfo {author} {\bibfnamefont {C.}~\bibnamefont
  {P\"ollmann}}, \bibinfo {author} {\bibfnamefont {P.}~\bibnamefont
  {Steinleitner}}, \bibinfo {author} {\bibfnamefont {U.}~\bibnamefont
  {Leierseder}}, \bibinfo {author} {\bibfnamefont {P.}~\bibnamefont {Nagler}},
  \bibinfo {author} {\bibfnamefont {G.}~\bibnamefont {Plechinger}}, \bibinfo
  {author} {\bibfnamefont {M.}~\bibnamefont {Porer}}, \bibinfo {author}
  {\bibfnamefont {R.}~\bibnamefont {Bratschitsch}}, \bibinfo {author}
  {\bibfnamefont {C.}~\bibnamefont {Sch\"uller}}, \bibinfo {author}
  {\bibfnamefont {T.}~\bibnamefont {Korn}},\ and\ \bibinfo {author}
  {\bibfnamefont {R.}~\bibnamefont {Huber}},\ }\href@noop {} {\bibfield
  {journal} {\bibinfo  {journal} {Nat. Mater.}\ }\textbf {\bibinfo {volume}
  {14}},\ \bibinfo {pages} {889} (\bibinfo {year} {2015})}\BibitemShut
  {NoStop}%
\bibitem [{\citenamefont {Zhang}\ \emph {et~al.}(2016)\citenamefont {Zhang},
  \citenamefont {Wang}, \citenamefont {Gao}, \citenamefont {Long},
  \citenamefont {Watson}, \citenamefont {Manfra}, \citenamefont {Belyanin},\
  and\ \citenamefont {Kono}}]{ZhangetAl16PRL}%
  \BibitemOpen
  \bibfield  {author} {\bibinfo {author} {\bibfnamefont {Q.}~\bibnamefont
  {Zhang}}, \bibinfo {author} {\bibfnamefont {Y.}~\bibnamefont {Wang}},
  \bibinfo {author} {\bibfnamefont {W.}~\bibnamefont {Gao}}, \bibinfo {author}
  {\bibfnamefont {Z.}~\bibnamefont {Long}}, \bibinfo {author} {\bibfnamefont
  {J.~D.}\ \bibnamefont {Watson}}, \bibinfo {author} {\bibfnamefont {M.~J.}\
  \bibnamefont {Manfra}}, \bibinfo {author} {\bibfnamefont {A.}~\bibnamefont
  {Belyanin}},\ and\ \bibinfo {author} {\bibfnamefont {J.}~\bibnamefont
  {Kono}},\ }\href@noop {} {\bibfield  {journal} {\bibinfo  {journal} {Phys.
  Rev. Lett.}\ }\textbf {\bibinfo {volume} {117}},\ \bibinfo {pages} {207402}
  (\bibinfo {year} {2016})}\BibitemShut {NoStop}%
\bibitem [{\citenamefont {Luo}\ \emph {et~al.}(2017)\citenamefont {Luo},
  \citenamefont {Men}, \citenamefont {Liu}, \citenamefont {Mudryk},
  \citenamefont {Zhao}, \citenamefont {Yao}, \citenamefont {Park},
  \citenamefont {Shinar}, \citenamefont {Shinar}, \citenamefont {Ho},
  \citenamefont {Perakis}, \citenamefont {Vela},\ and\ \citenamefont
  {Wang}}]{LuoetAl17NC}%
  \BibitemOpen
  \bibfield  {author} {\bibinfo {author} {\bibfnamefont {L.}~\bibnamefont
  {Luo}}, \bibinfo {author} {\bibfnamefont {L.}~\bibnamefont {Men}}, \bibinfo
  {author} {\bibfnamefont {Z.}~\bibnamefont {Liu}}, \bibinfo {author}
  {\bibfnamefont {Y.}~\bibnamefont {Mudryk}}, \bibinfo {author} {\bibfnamefont
  {X.}~\bibnamefont {Zhao}}, \bibinfo {author} {\bibfnamefont {Y.}~\bibnamefont
  {Yao}}, \bibinfo {author} {\bibfnamefont {J.~M.}\ \bibnamefont {Park}},
  \bibinfo {author} {\bibfnamefont {R.}~\bibnamefont {Shinar}}, \bibinfo
  {author} {\bibfnamefont {J.}~\bibnamefont {Shinar}}, \bibinfo {author}
  {\bibfnamefont {K.-M.}\ \bibnamefont {Ho}}, \bibinfo {author} {\bibfnamefont
  {I.~E.}\ \bibnamefont {Perakis}}, \bibinfo {author} {\bibfnamefont
  {J.}~\bibnamefont {Vela}},\ and\ \bibinfo {author} {\bibfnamefont
  {J.}~\bibnamefont {Wang}},\ }\href@noop {} {\bibfield  {journal} {\bibinfo
  {journal} {Nat. Commun.}\ }\textbf {\bibinfo {volume} {8}},\ \bibinfo {pages}
  {15565} (\bibinfo {year} {2017})}\BibitemShut {NoStop}%
\bibitem [{\citenamefont {Luo}\ \emph {et~al.}(2019)\citenamefont {Luo},
  \citenamefont {Liu}, \citenamefont {Yang}, \citenamefont {Vaswani},
  \citenamefont {Cheng}, \citenamefont {Park},\ and\ \citenamefont
  {Wang}}]{LuoetAl19PRM}%
  \BibitemOpen
  \bibfield  {author} {\bibinfo {author} {\bibfnamefont {L.}~\bibnamefont
  {Luo}}, \bibinfo {author} {\bibfnamefont {Z.}~\bibnamefont {Liu}}, \bibinfo
  {author} {\bibfnamefont {X.}~\bibnamefont {Yang}}, \bibinfo {author}
  {\bibfnamefont {C.}~\bibnamefont {Vaswani}}, \bibinfo {author} {\bibfnamefont
  {D.}~\bibnamefont {Cheng}}, \bibinfo {author} {\bibfnamefont {J.-M.}\
  \bibnamefont {Park}},\ and\ \bibinfo {author} {\bibfnamefont
  {J.}~\bibnamefont {Wang}},\ }\href@noop {} {\bibfield  {journal} {\bibinfo
  {journal} {Phys. Rev. Materials}\ }\textbf {\bibinfo {volume} {3}},\ \bibinfo
  {pages} {026003} (\bibinfo {year} {2019})}\BibitemShut {NoStop}%
\bibitem [{\citenamefont {Sekiguchi}\ and\ \citenamefont
  {Shimano}(2015)}]{SekiguchiShimano15PRB}%
  \BibitemOpen
  \bibfield  {author} {\bibinfo {author} {\bibfnamefont {F.}~\bibnamefont
  {Sekiguchi}}\ and\ \bibinfo {author} {\bibfnamefont {R.}~\bibnamefont
  {Shimano}},\ }\href@noop {} {\bibfield  {journal} {\bibinfo  {journal} {Phys.
  Rev. B}\ }\textbf {\bibinfo {volume} {91}},\ \bibinfo {pages} {155202}
  (\bibinfo {year} {2015})}\BibitemShut {NoStop}%
\bibitem [{\citenamefont {MacDonald}\ and\ \citenamefont
  {Rezayi}(1990)}]{MacDonaldRezayi90PRB}%
  \BibitemOpen
  \bibfield  {author} {\bibinfo {author} {\bibfnamefont {A.~H.}\ \bibnamefont
  {MacDonald}}\ and\ \bibinfo {author} {\bibfnamefont {E.~H.}\ \bibnamefont
  {Rezayi}},\ }\href@noop {} {\bibfield  {journal} {\bibinfo  {journal} {Phys.
  Rev. B}\ }\textbf {\bibinfo {volume} {42}},\ \bibinfo {pages} {3224}
  (\bibinfo {year} {1990})}\BibitemShut {NoStop}%
\bibitem [{\citenamefont {Dzyubenko}\ and\ \citenamefont
  {Lozovik}(1991)}]{DzyubenkoLozovik91JPA}%
  \BibitemOpen
  \bibfield  {author} {\bibinfo {author} {\bibfnamefont {A.~B.}\ \bibnamefont
  {Dzyubenko}}\ and\ \bibinfo {author} {\bibfnamefont {Y.~E.}\ \bibnamefont
  {Lozovik}},\ }\href@noop {} {\bibfield  {journal} {\bibinfo  {journal} {J.
  Phys. A: Math. Gen.}\ }\textbf {\bibinfo {volume} {24}},\ \bibinfo {pages}
  {415} (\bibinfo {year} {1991})}\BibitemShut {NoStop}%
\bibitem [{\citenamefont {Apal'kov}\ and\ \citenamefont
  {Rashba}(1991)}]{ApalkonRashba91JETP}%
  \BibitemOpen
  \bibfield  {author} {\bibinfo {author} {\bibfnamefont {V.~M.}\ \bibnamefont
  {Apal'kov}}\ and\ \bibinfo {author} {\bibfnamefont {E.~I.}\ \bibnamefont
  {Rashba}},\ }\href@noop {} {\bibfield  {journal} {\bibinfo  {journal} {JETP
  Lett.}\ }\textbf {\bibinfo {volume} {53}},\ \bibinfo {pages} {442} (\bibinfo
  {year} {1991})}\BibitemShut {NoStop}%
\bibitem [{\citenamefont {MacDonald}\ \emph {et~al.}(1992)\citenamefont
  {MacDonald}, \citenamefont {Rezayi},\ and\ \citenamefont
  {Keller}}]{MacDonaldetAl92PRL}%
  \BibitemOpen
  \bibfield  {author} {\bibinfo {author} {\bibfnamefont {A.~H.}\ \bibnamefont
  {MacDonald}}, \bibinfo {author} {\bibfnamefont {E.~H.}\ \bibnamefont
  {Rezayi}},\ and\ \bibinfo {author} {\bibfnamefont {D.}~\bibnamefont
  {Keller}},\ }\href@noop {} {\bibfield  {journal} {\bibinfo  {journal} {Phys.
  Rev. Lett.}\ }\textbf {\bibinfo {volume} {68}},\ \bibinfo {pages} {1939}
  (\bibinfo {year} {1992})}\BibitemShut {NoStop}%
\bibitem [{\citenamefont {Yoon}\ \emph {et~al.}(1997)\citenamefont {Yoon},
  \citenamefont {Sturge},\ and\ \citenamefont {Pfeiffer}}]{YoonetAl97SSC}%
  \BibitemOpen
  \bibfield  {author} {\bibinfo {author} {\bibfnamefont {H.~W.}\ \bibnamefont
  {Yoon}}, \bibinfo {author} {\bibfnamefont {M.~D.}\ \bibnamefont {Sturge}},\
  and\ \bibinfo {author} {\bibfnamefont {L.~N.}\ \bibnamefont {Pfeiffer}},\
  }\href@noop {} {\bibfield  {journal} {\bibinfo  {journal} {Solid State
  Commun.}\ }\textbf {\bibinfo {volume} {104}},\ \bibinfo {pages} {287}
  (\bibinfo {year} {1997})}\BibitemShut {NoStop}%
\bibitem [{\citenamefont {Rashba}\ \emph {et~al.}(2000)\citenamefont {Rashba},
  \citenamefont {Sturge}, \citenamefont {Yoon},\ and\ \citenamefont
  {Pfeiffer}}]{RashbaetAl00SSC}%
  \BibitemOpen
  \bibfield  {author} {\bibinfo {author} {\bibfnamefont {E.~I.}\ \bibnamefont
  {Rashba}}, \bibinfo {author} {\bibfnamefont {M.~D.}\ \bibnamefont {Sturge}},
  \bibinfo {author} {\bibfnamefont {H.~W.}\ \bibnamefont {Yoon}},\ and\
  \bibinfo {author} {\bibfnamefont {L.~N.}\ \bibnamefont {Pfeiffer}},\
  }\href@noop {} {\bibfield  {journal} {\bibinfo  {journal} {Solid State
  Commun.}\ }\textbf {\bibinfo {volume} {114}},\ \bibinfo {pages} {593}
  (\bibinfo {year} {2000})}\BibitemShut {NoStop}%
\bibitem [{\citenamefont {Kim}\ \emph {et~al.}(2013)\citenamefont {Kim},
  \citenamefont {Lee}, \citenamefont {Noe}, \citenamefont {Wang}, \citenamefont
  {W\'ojcik}, \citenamefont {McGill}, \citenamefont {Reitze}, \citenamefont
  {Belyanin},\ and\ \citenamefont {Kono}}]{KimetAl13PRB}%
  \BibitemOpen
  \bibfield  {author} {\bibinfo {author} {\bibfnamefont {J.-H.}\ \bibnamefont
  {Kim}}, \bibinfo {author} {\bibfnamefont {J.}~\bibnamefont {Lee}}, \bibinfo
  {author} {\bibfnamefont {G.~T.}\ \bibnamefont {Noe}}, \bibinfo {author}
  {\bibfnamefont {Y.}~\bibnamefont {Wang}}, \bibinfo {author} {\bibfnamefont
  {A.~K.}\ \bibnamefont {W\'ojcik}}, \bibinfo {author} {\bibfnamefont {S.~A.}\
  \bibnamefont {McGill}}, \bibinfo {author} {\bibfnamefont {D.~H.}\
  \bibnamefont {Reitze}}, \bibinfo {author} {\bibfnamefont {A.~A.}\
  \bibnamefont {Belyanin}},\ and\ \bibinfo {author} {\bibfnamefont
  {J.}~\bibnamefont {Kono}},\ }\href@noop {} {\bibfield  {journal} {\bibinfo
  {journal} {Phys. Rev. B}\ }\textbf {\bibinfo {volume} {87}},\ \bibinfo
  {pages} {045304} (\bibinfo {year} {2013})}\BibitemShut {NoStop}%
\bibitem [{\citenamefont {Hayamizu}\ \emph {et~al.}(2007)\citenamefont
  {Hayamizu}, \citenamefont {Yoshita}, \citenamefont {Takahashi}, \citenamefont
  {Akiyama}, \citenamefont {Ning}, \citenamefont {Pfeiffer},\ and\
  \citenamefont {West}}]{HayamizuetAl07PRL}%
  \BibitemOpen
  \bibfield  {author} {\bibinfo {author} {\bibfnamefont {Y.}~\bibnamefont
  {Hayamizu}}, \bibinfo {author} {\bibfnamefont {M.}~\bibnamefont {Yoshita}},
  \bibinfo {author} {\bibfnamefont {Y.}~\bibnamefont {Takahashi}}, \bibinfo
  {author} {\bibfnamefont {H.}~\bibnamefont {Akiyama}}, \bibinfo {author}
  {\bibfnamefont {C.~Z.}\ \bibnamefont {Ning}}, \bibinfo {author}
  {\bibfnamefont {L.~N.}\ \bibnamefont {Pfeiffer}},\ and\ \bibinfo {author}
  {\bibfnamefont {K.~W.}\ \bibnamefont {West}},\ }\href@noop {} {\bibfield
  {journal} {\bibinfo  {journal} {Phys. Rev. Lett.}\ }\textbf {\bibinfo
  {volume} {99}},\ \bibinfo {pages} {167403} (\bibinfo {year}
  {2007})}\BibitemShut {NoStop}%
\bibitem [{\citenamefont {Kawada}\ \emph {et~al.}(2014)\citenamefont {Kawada},
  \citenamefont {Yasuda}, \citenamefont {Nakanishi}, \citenamefont {Akiyama},
  \citenamefont {Hakamata},\ and\ \citenamefont {Takahashi}}]{KawadaetAl14OL}%
  \BibitemOpen
  \bibfield  {author} {\bibinfo {author} {\bibfnamefont {Y.}~\bibnamefont
  {Kawada}}, \bibinfo {author} {\bibfnamefont {T.}~\bibnamefont {Yasuda}},
  \bibinfo {author} {\bibfnamefont {A.}~\bibnamefont {Nakanishi}}, \bibinfo
  {author} {\bibfnamefont {K.}~\bibnamefont {Akiyama}}, \bibinfo {author}
  {\bibfnamefont {K.}~\bibnamefont {Hakamata}},\ and\ \bibinfo {author}
  {\bibfnamefont {H.}~\bibnamefont {Takahashi}},\ }\href@noop {} {\bibfield
  {journal} {\bibinfo  {journal} {Opt. Lett.}\ }\textbf {\bibinfo {volume}
  {39}},\ \bibinfo {pages} {2794} (\bibinfo {year} {2014})}\BibitemShut
  {NoStop}%
\bibitem [{\citenamefont {Li}\ \emph {et~al.}(2018)\citenamefont {Li},
  \citenamefont {Bamba}, \citenamefont {Zhang}, \citenamefont {Fallahi},
  \citenamefont {Gardner}, \citenamefont {Gao}, \citenamefont {Lou},
  \citenamefont {Yoshioka}, \citenamefont {Manfra},\ and\ \citenamefont
  {Kono}}]{LietAl18NP}%
  \BibitemOpen
  \bibfield  {author} {\bibinfo {author} {\bibfnamefont {X.}~\bibnamefont
  {Li}}, \bibinfo {author} {\bibfnamefont {M.}~\bibnamefont {Bamba}}, \bibinfo
  {author} {\bibfnamefont {Q.}~\bibnamefont {Zhang}}, \bibinfo {author}
  {\bibfnamefont {S.}~\bibnamefont {Fallahi}}, \bibinfo {author} {\bibfnamefont
  {G.~C.}\ \bibnamefont {Gardner}}, \bibinfo {author} {\bibfnamefont
  {W.}~\bibnamefont {Gao}}, \bibinfo {author} {\bibfnamefont {M.}~\bibnamefont
  {Lou}}, \bibinfo {author} {\bibfnamefont {K.}~\bibnamefont {Yoshioka}},
  \bibinfo {author} {\bibfnamefont {M.~J.}\ \bibnamefont {Manfra}},\ and\
  \bibinfo {author} {\bibfnamefont {J.}~\bibnamefont {Kono}},\ }\href@noop {}
  {\bibfield  {journal} {\bibinfo  {journal} {Nat. Photon.}\ }\textbf {\bibinfo
  {volume} {12}},\ \bibinfo {pages} {324} (\bibinfo {year} {2018})}\BibitemShut
  {NoStop}%
\bibitem [{SM()}]{SM}%
  \BibitemOpen
  \bibinfo {note} {See Supplemental Material at [URL will be inserted by
  publisher] for additional details, which includes Refs. [41-46].}\BibitemShut {Stop}%
\bibitem [{\citenamefont {Kira}\ and\ \citenamefont
  {Koch}(2004)}]{KiraKoch04PRL}%
  \BibitemOpen
  \bibfield  {author} {\bibinfo {author} {\bibfnamefont {M.}~\bibnamefont
  {Kira}}\ and\ \bibinfo {author} {\bibfnamefont {S.~W.}\ \bibnamefont
  {Koch}},\ }\href@noop {} {\bibfield  {journal} {\bibinfo  {journal} {Phys.
  Rev. Lett.}\ }\textbf {\bibinfo {volume} {93}},\ \bibinfo {pages} {076402}
  (\bibinfo {year} {2004})}\BibitemShut {NoStop}%
\bibitem [{\citenamefont {Harley}\ and\ \citenamefont
  {Snelling}(1996)}]{HarleyetAL96PRB}%
  \BibitemOpen
  \bibfield  {author} {\bibinfo {author} {\bibfnamefont {R.~T.}\ \bibnamefont
  {Harley}}\ and\ \bibinfo {author} {\bibfnamefont {M.~J.}\ \bibnamefont
  {Snelling}},\ }\href {https://doi.org/10.1103/PhysRevB.53.9561} {\bibfield
  {journal} {\bibinfo  {journal} {Phys. Rev. B}\ }\textbf {\bibinfo {volume}
  {53}},\ \bibinfo {pages} {9561} (\bibinfo {year} {1996})}\BibitemShut
  {NoStop}%
\bibitem [{\citenamefont {Zhukavin}\ \emph {et~al.}(2007)\citenamefont
  {Zhukavin}, \citenamefont {Shastin}, \citenamefont {Pavlov}, \citenamefont
  {Hübers}, \citenamefont {Hovenier}, \citenamefont {Klaassen},\ and\
  \citenamefont {van~der Meer}}]{ZhukavinetAL07JAP}%
  \BibitemOpen
  \bibfield  {author} {\bibinfo {author} {\bibfnamefont {R.~K.}\ \bibnamefont
  {Zhukavin}}, \bibinfo {author} {\bibfnamefont {V.~N.}\ \bibnamefont
  {Shastin}}, \bibinfo {author} {\bibfnamefont {S.~G.}\ \bibnamefont {Pavlov}},
  \bibinfo {author} {\bibfnamefont {H.-W.}\ \bibnamefont {Hübers}}, \bibinfo
  {author} {\bibfnamefont {J.~N.}\ \bibnamefont {Hovenier}}, \bibinfo {author}
  {\bibfnamefont {T.~O.}\ \bibnamefont {Klaassen}},\ and\ \bibinfo {author}
  {\bibfnamefont {A.~F.~G.}\ \bibnamefont {van~der Meer}},\ }\href
  {https://doi.org/10.1063/1.2804756} {\bibfield  {journal} {\bibinfo
  {journal} {J. Appl. Phys.}\ }\textbf {\bibinfo {volume} {102}},\ \bibinfo
  {pages} {093104} (\bibinfo {year} {2007})}\BibitemShut {NoStop}%
\bibitem [{\citenamefont {Zakhar’in}\ \emph {et~al.}(2010)\citenamefont
  {Zakhar’in}, \citenamefont {Andrianov}, \citenamefont {Egorov},\ and\
  \citenamefont {Zinov’ev}}]{ZakharinetAL10APL}%
  \BibitemOpen
  \bibfield  {author} {\bibinfo {author} {\bibfnamefont {A.~O.}\ \bibnamefont
  {Zakhar’in}}, \bibinfo {author} {\bibfnamefont {A.~V.}\ \bibnamefont
  {Andrianov}}, \bibinfo {author} {\bibfnamefont {A.~Y.}\ \bibnamefont
  {Egorov}},\ and\ \bibinfo {author} {\bibfnamefont {N.~N.}\ \bibnamefont
  {Zinov’ev}},\ }\href {https://doi.org/10.1063/1.3441401} {\bibfield
  {journal} {\bibinfo  {journal} {Appl. Phys. Lett.}\ }\textbf {\bibinfo
  {volume} {96}},\ \bibinfo {pages} {211118} (\bibinfo {year}
  {2010})}\BibitemShut {NoStop}%
\bibitem [{\citenamefont {Makhov}\ \emph
  {et~al.}(2019{\natexlab{a}})\citenamefont {Makhov}, \citenamefont {Panevin},
  \citenamefont {Firsov}, \citenamefont {Vorobjev}, \citenamefont {Vasil'ev},\
  and\ \citenamefont {Maleev}}]{MakhovetAL19JLum}%
  \BibitemOpen
  \bibfield  {author} {\bibinfo {author} {\bibfnamefont {I.~S.}\ \bibnamefont
  {Makhov}}, \bibinfo {author} {\bibfnamefont {V.~Y.}\ \bibnamefont {Panevin}},
  \bibinfo {author} {\bibfnamefont {D.~A.}\ \bibnamefont {Firsov}}, \bibinfo
  {author} {\bibfnamefont {L.~E.}\ \bibnamefont {Vorobjev}}, \bibinfo {author}
  {\bibfnamefont {A.~P.}\ \bibnamefont {Vasil'ev}},\ and\ \bibinfo {author}
  {\bibfnamefont {N.~A.}\ \bibnamefont {Maleev}},\ }\href
  {http://www.sciencedirect.com/science/article/pii/S0022231318318544}
  {\bibfield  {journal} {\bibinfo  {journal} {J. Lumin.}\ }\textbf {\bibinfo
  {volume} {210}},\ \bibinfo {pages} {352 } (\bibinfo {year}
  {2019}{\natexlab{a}})}\BibitemShut {NoStop}%
\bibitem [{\citenamefont {Makhov}\ \emph
  {et~al.}(2019{\natexlab{b}})\citenamefont {Makhov}, \citenamefont {Panevin},
  \citenamefont {Firsov}, \citenamefont {Vorobjev},\ and\ \citenamefont
  {Klimko}}]{MakhovetAL19JAP}%
  \BibitemOpen
  \bibfield  {author} {\bibinfo {author} {\bibfnamefont {I.~S.}\ \bibnamefont
  {Makhov}}, \bibinfo {author} {\bibfnamefont {V.~Y.}\ \bibnamefont {Panevin}},
  \bibinfo {author} {\bibfnamefont {D.~A.}\ \bibnamefont {Firsov}}, \bibinfo
  {author} {\bibfnamefont {L.~E.}\ \bibnamefont {Vorobjev}},\ and\ \bibinfo
  {author} {\bibfnamefont {G.~V.}\ \bibnamefont {Klimko}},\ }\href
  {https://doi.org/10.1063/1.5121835} {\bibfield  {journal} {\bibinfo
  {journal} {J. Appl. Phys.}\ }\textbf {\bibinfo {volume} {126}},\ \bibinfo
  {pages} {175702} (\bibinfo {year} {2019}{\natexlab{b}})}\BibitemShut
  {NoStop}%
\bibitem [{\citenamefont {Snelling}\ \emph {et~al.}(1992)\citenamefont
  {Snelling}, \citenamefont {Blackwood}, \citenamefont {McDonagh},
  \citenamefont {Harley},\ and\ \citenamefont {Foxon}}]{SnellingetAl92PRB}%
  \BibitemOpen
  \bibfield  {author} {\bibinfo {author} {\bibfnamefont {M.~J.}\ \bibnamefont
  {Snelling}}, \bibinfo {author} {\bibfnamefont {E.}~\bibnamefont {Blackwood}},
  \bibinfo {author} {\bibfnamefont {C.~J.}\ \bibnamefont {McDonagh}}, \bibinfo
  {author} {\bibfnamefont {R.~T.}\ \bibnamefont {Harley}},\ and\ \bibinfo
  {author} {\bibfnamefont {C.~T.~B.}\ \bibnamefont {Foxon}},\ }\href@noop {}
  {\bibfield  {journal} {\bibinfo  {journal} {Phys. Rev. B}\ }\textbf {\bibinfo
  {volume} {45}},\ \bibinfo {pages} {3922} (\bibinfo {year}
  {1992})}\BibitemShut {NoStop}%
\bibitem [{\citenamefont {Traynor}\ \emph {et~al.}(1997)\citenamefont
  {Traynor}, \citenamefont {Warburton}, \citenamefont {Snelling},\ and\
  \citenamefont {Harley}}]{TraynoretAl97PRB}%
  \BibitemOpen
  \bibfield  {author} {\bibinfo {author} {\bibfnamefont {N.~J.}\ \bibnamefont
  {Traynor}}, \bibinfo {author} {\bibfnamefont {R.~J.}\ \bibnamefont
  {Warburton}}, \bibinfo {author} {\bibfnamefont {M.~J.}\ \bibnamefont
  {Snelling}},\ and\ \bibinfo {author} {\bibfnamefont {R.~T.}\ \bibnamefont
  {Harley}},\ }\href@noop {} {\bibfield  {journal} {\bibinfo  {journal} {Phys.
  Rev. B}\ }\textbf {\bibinfo {volume} {55}},\ \bibinfo {pages} {15701}
  (\bibinfo {year} {1997})}\BibitemShut {NoStop}%
\bibitem [{\citenamefont {Kioseoglou}\ \emph {et~al.}(2000)\citenamefont
	{Kioseoglou}, \citenamefont {Cheong}, \citenamefont {Nickel}, \citenamefont
	{Petrou}, \citenamefont {McCombe},\ and\ \citenamefont
	{Schaff}}]{Kioseoglou2000}%
\BibitemOpen
\bibfield  {author} {\bibinfo {author} {\bibfnamefont {G.}~\bibnamefont
		{Kioseoglou}}, \bibinfo {author} {\bibfnamefont {H.~D.}\ \bibnamefont
		{Cheong}}, \bibinfo {author} {\bibfnamefont {H.~A.}\ \bibnamefont {Nickel}},
	\bibinfo {author} {\bibfnamefont {A.}~\bibnamefont {Petrou}}, \bibinfo
	{author} {\bibfnamefont {B.~D.}\ \bibnamefont {McCombe}},\ and\ \bibinfo
	{author} {\bibfnamefont {W.}~\bibnamefont {Schaff}},\ }\href
{https://doi.org/10.1103/PhysRevB.61.4780} {\bibfield  {journal} {\bibinfo
		{journal} {Phys. Rev. B}\ }\textbf {\bibinfo {volume} {61}},\ \bibinfo
	{pages} {4780} (\bibinfo {year} {2000})}\BibitemShut {NoStop}%
\bibitem [{\citenamefont {Shanabrook}\ \emph {et~al.}(1987)\citenamefont
	{Shanabrook}, \citenamefont {Glembocki},\ and\ \citenamefont
	{Beard}}]{Shanabrook1987}%
\BibitemOpen
\bibfield  {author} {\bibinfo {author} {\bibfnamefont {B.~V.}\ \bibnamefont
		{Shanabrook}}, \bibinfo {author} {\bibfnamefont {O.~J.}\ \bibnamefont
		{Glembocki}},\ and\ \bibinfo {author} {\bibfnamefont {W.~T.}\ \bibnamefont
		{Beard}},\ }\href {https://doi.org/10.1103/PhysRevB.35.2540} {\bibfield
	{journal} {\bibinfo  {journal} {Phys. Rev. B}\ }\textbf {\bibinfo {volume}
		{35}},\ \bibinfo {pages} {2540} (\bibinfo {year} {1987})}\BibitemShut
{NoStop}%
\bibitem [{\citenamefont {Phillips}\ \emph {et~al.}(1992)\citenamefont
	{Phillips}, \citenamefont {Lovering}, \citenamefont {Denton},\ and\
	\citenamefont {Smith}}]{Phillips1992}%
\BibitemOpen
\bibfield  {author} {\bibinfo {author} {\bibfnamefont {R.~T.}\ \bibnamefont
		{Phillips}}, \bibinfo {author} {\bibfnamefont {D.~J.}\ \bibnamefont
		{Lovering}}, \bibinfo {author} {\bibfnamefont {G.~J.}\ \bibnamefont
		{Denton}},\ and\ \bibinfo {author} {\bibfnamefont {G.~W.}\ \bibnamefont
		{Smith}},\ }\href {https://doi.org/10.1103/PhysRevB.45.4308} {\bibfield
	{journal} {\bibinfo  {journal} {Phys. Rev. B}\ }\textbf {\bibinfo {volume}
		{45}},\ \bibinfo {pages} {4308} (\bibinfo {year} {1992})}\BibitemShut
{NoStop}%
\bibitem [{\citenamefont {Nuss}\ and\ \citenamefont
	{Orenstein}(1998)}]{Nuss1998}%
\BibitemOpen
\bibfield  {author} {\bibinfo {author} {\bibfnamefont {M.~C.}\ \bibnamefont
		{Nuss}}\ and\ \bibinfo {author} {\bibfnamefont {J.}~\bibnamefont
		{Orenstein}},\ }\bibinfo {title} {Terahertz time-domain spectroscopy},\ in\
\href {https://doi.org/10.1007/BFb0103419} {\emph {\bibinfo {booktitle}
		{Millimeter and Submillimeter Wave Spectroscopy of Solids}}},\ \bibinfo
{editor} {edited by\ \bibinfo {editor} {\bibfnamefont {G.}~\bibnamefont
		{Gr{\"u}ner}}}\ (\bibinfo  {publisher} {Springer Berlin Heidelberg},\
\bibinfo {address} {Berlin, Heidelberg},\ \bibinfo {year} {1998})\ pp.\
\bibinfo {pages} {7--50}\BibitemShut {NoStop}%
\end{thebibliography}

\end{document}